\documentclass[twocolumn,amssymb,floatfix,superscriptaddress,nofootinbib]{revtex4}
\usepackage{amsmath,amssymb,color}
\usepackage{graphicx}

\newcommand{\be}{\begin{eqnarray}}
\newcommand{\ee}{\end{eqnarray}}
\def\ben{\begin{equation}}
\def\een{\end{equation}}
\def\bena{\begin{eqnarray}}
\def\eena{\end{eqnarray}}

\newcommand{\Sarah}{\color{black}}

\newcommand{\ha}{\phi}

\begin{document}

\title{Axion Dark Matter and Planck favor non-minimal couplings to gravity}

\author{Sarah Folkerts}
\email{sarah.folkerts@lmu.de}
\affiliation{Arnold Sommerfeld Center, Ludwig-Maximilians-University, Theresienstr. 37, 80333 M\"unchen, Germany}
\author{Cristiano Germani}
\email{cristiano.germani@lmu.de}
\affiliation{Arnold Sommerfeld Center, Ludwig-Maximilians-University, Theresienstr. 37, 80333 M\"unchen, Germany}
\author{Javier Redondo}
\email{javier.redondo@lmu.de}
\affiliation{Arnold Sommerfeld Center, Ludwig-Maximilians-University, Theresienstr. 37, 80333 M\"unchen, Germany}
\affiliation{Max-Planck-Institut f\"ur Physik (Werner-Heisenberg-Institut) , F\"ohringer Ring 6, 80805 M\"unchen,
Germany} 

\begin{abstract}
Constraints on inflationary scenarios and isocurvature perturbations have excluded the simplest and most generic models of dark matter based on QCD axions. Considering non-minimal kinetic couplings of scalar fields to gravity substantially changes this picture. 
The axion can account for the observed dark matter density avoiding the overproduction of isocurvature fluctuations.
Finally, we show that assuming the same non-minimal kinetic coupling to the axion (dark matter) and to the standard model Higgs boson (inflaton) provides a minimal picture of  early time cosmology.

\end{abstract}

\pacs{}

\maketitle
\section*{Introduction}~\label{sec0}

 The results of Planck are striking for cosmology~\cite{planck}. The absence of non-Gaussianities generically point to inflationary models where the anisotropies of the Cosmic Microwave Background (CMB) are effectively generated by a single scalar field during inflation~\cite{Ade:2013ydc}. This is due to the fact that large non-Gaussianities are generically produced by any non-negligible coupling between evolving light scalars during inflation.

The fact that no gravitational waves have been observed puts a limit on the Hubble scale during inflation $H_I<9\times 10^{13}$ GeV~\cite{planck} independently of the model of inflation. This, together with the measured power of temperature fluctuations $P_s\approx 2.2\times 10^{-9}$ and their spectral tilt $n_s\approx 0.96$, severely constrains single-field inflation. 

Large-field scenarios (chaotic slow-roll inflation)~\cite{linde} require typical inflationary scales $H_I\sim {\cal O}(10^{13}{\rm GeV})$ and are disfavored with respect to models with a non-minimal kinetic coupling to gravity~\cite{yuki}. This is because the ratio of tensor to scalar perturbations depends on ``how fast'' the inflaton moves in its potential. In non-minimal models the inflaton evolves slower than the minimally coupled cousins for a fixed $n_s$, thus relaxing the tension~\cite{sloth}.    

In slow-roll small field inflation, e.g. natural inflation models~\cite{natural}, $n_s$ receives a contribution from a tachyonic mass which allows smaller $H_I$ within the bound from gravitational waves. However, this is at the expense of introducing unreliable trans-Planckian energy scales~\cite{gia}. 
Once again, non-minimal couplings to gravity can evade this problem~\cite{uv}.

Finally, if the CMB anisotropies are not generated by the inflaton itself, but are transferred from isocurvature-type to adiabatic by the decay of a light scalar field (the curvaton~\cite{Mollerach:1989hu}), constraints on the absence of non-Gaussianities and residual isocurvature perturbations can be avoided~\cite{Fonseca:2011aa}. In this case however, it is hard to gain any information about the nature of the inflaton with CMB physics \footnote{Unless constraints on non-Gaussianities   improve notably in the future.}.

The implications of Planck transcend the physics of inflation and have far-reaching consequences for other aspects of cosmology, like the nature of dark matter (DM). 

The ``invisible'' QCD axion is one of the better motivated DM candidates~\cite{Visinelli:2009zm,Wantz:2009it,Sikivie:2006ni}. 
It is a hypothetical $0^-$ particle with a very small mass ($m_a\lesssim 10$ meV) that appears~\cite{Weinberg:1977ma,Wilczek:1977pj} as a consequence of the Peccei-Quinn solution to the strong CP problem of the SM~\cite{Peccei:1977hh,Peccei:2006as}. It requires peculiar laboratory searches~\cite{Sikivie:1983ip,Asztalos:2009yp,Horns:2012jf} and might have distinctive signatures in structure formation~\cite{Erken:2011dz}.

The axion can be described by a model-independent effective-field-theory up to a strong coupling scale $f_a$, see~\cite{Georgi:1986df}. The axion would be a massless field if it were not for its coupling to the QCD Chern-Simons form (${\cal L }\ni  G^a_{\mu\nu}\widetilde G^{a\mu\nu} a/f_a$), see e.g. III.B of~\cite{Kim:2008hd}. 
QCD instantons induce a potential for the axion that is periodic $a\to a+2\pi f_a$. One usually parameterizes the axion with an angle $\Theta\in [-\pi,\pi]$ defining $a=\Theta f_a$. 
The potential and thus the axion mass is strongly suppressed at temperatures larger than the QCD scale, $T>\Lambda_{\rm QCD}$. 
This description can be used during inflation if the inflationary scale $H_I$ is below $f_a$, the scenario we focus on in this letter. 

Before inflation, the axion vacuum expectation value (vev) settles to different values within each causally disconnected region since its potential is flat. 
During inflation a tiny causal region inflates to host our universe, thus making the axion vev ($a_i=\Theta_i f_a$) homogeneous. After inflation, the universe reheats and cools down. Around $T\sim$ GeV the axion mass is generated, and the axion field starts oscillating around the minimum of its potential producing DM particles with a density proportional to $\Theta_i^2$. This ``misalignment mechanism''~\cite{Preskill:1982cy,Abbott:1982af,Dine:1982ah} produces a cold form of DM. 
The observed amount of DM can be obtained for a broad range of values of $f_a>10^{10}$ GeV~\cite{Wantz:2009it}. The required value $\Theta_i$ can be supported by anthropic arguments~\cite{Preskill:1982cy,Pi:1984pv}. For this reason this scenario is sometimes called the ``anthropic axion scenario''. 
Experimental searches for axion DM are sensitive for $f_a\sim 10^{12}$ GeV ($\mu$eV masses)~\cite{Asztalos:2009yp} and new ideas exist to explore much larger values~\cite{Graham:2011qk,Graham:2013gfa,Budker:2013hfa}.

The anthropic axion DM paradigm has been severely constrained by cosmology but, after Planck, it is more than ever.  
During inflation, the massless axion has quantum fluctuations around $a_i$. These induce DM isocurvature fluctuations of order $\langle\delta \Theta^2\rangle/\Theta^2_i\sim H_I^2/\Theta^2_i f_a^2$ when the axion mass sets in. 
They modify the temperature power spectrum of the CMB by shifting the acoustic oscillations towards smaller scales~\cite{Lyth:1989pb} and have not been observed by Planck, implying a tight constraint $\langle\delta \Theta^2\rangle/\Theta^2_i<8.6\times 10^{-11}$.  

Even for $\Theta_i \sim \pi$,  Planck constraints require $H_I\ll f_a$. Plugging the numbers (as shown later on), only two possible solutions of the isocurvature problem turn out to be compatible with the right relic abundance of DM: $f_a>M_p$, or $H_I<10^{10}$ GeV.
The first option is theoretically unreliable since it invokes trans-Planckian physics~\cite{gia}. 
As for the second, in the minimally coupled case, large-field inflation requires $H_I\sim 10^{13}$ GeV and small-field inflation needs trans-Planckian scales, so the only apparent solution is to introduce new degrees of freedom, as in the case of the curvaton. 


In this letter, we show that considering non-minimal kinetic couplings to gravity allows for two alternative solutions to the isocurvature problem {\it without introducing new degrees of freedom}. 

As the first alternative, we consider the natural inflationary scenario of \cite{uv}. There, a consistent single-field scenario, with sufficiently low Hubble scale $H_I$ so to fulfill the isocurvature constraints of Planck, is obtained by a non-minimal derivative coupling of an hidden axion to curvatures.

The second alternative invokes instead a non-minimal derivative coupling of the QCD axion to gravity. The new coupling does not affect the density of DM, but can suppress the isocurvature perturbations during inflation.\footnote{This mechanism can be invoked to evade the isocurvature constraints of other low-mass DM fields~\cite{Arias:2012az} equally well.}

  
  As an interesting possibility, we will present a minimalistic cosmological scenario in which the inflaton is the standard model (SM) Higgs boson and the QCD axion accounts for the observed DM. Both scalar fields require non-minimal couplings to gravity, the Higgs to fit Planck data and the axion to avoid isocurvature constraints. Remarkably, the required non-minimal couplings can be of similar order of magnitude. 

Finally, we would like to mention that alternative scenarios to the anthropic axion are also severely constrained by cosmology. 
The axion can be modeled as a Nambu-Goldstone boson appearing only at temperatures  below the so-called Peccei-Quinn phase-transition ($\sim f_a$). If this transition happens after inflation, isocurvature perturbations are not generated simply because the axion does not exist yet. However, during this transition topological defects are produced, which in the most generic cases over-close the Universe and are excluded~\cite{Sikivie:1982qv,Hiramatsu:2010yn}. 


\section{Non-minimal couplings during Inflation}\label{sec1} 

The paradigm of inflation is based on a massless spin-2 particle (the graviton)  interacting with a spin-0 particle (the inflaton), i.e. three degrees of freedom. Since gravity is non-renormalizable, we are allowed to consider all possible interactions that do not change the number of degrees of freedom. As we are only interested in small derivative expansions (slow-roll), we consider Lagrangians up to two derivatives in the (canonical) scalar. Couplings of type $U(\phi)R$ (see e.g. \cite{higgs}) are equivalent to a redefinition of the scalar potential $V(\phi)$ via a conformal transformation of the metric, the so-called ``Einstein frame". Therefore, the most generic ghost-free action in Einstein frame, linear in curvatures, reads~\cite{new}
\be\label{L}
\frac{1}{2}\int \!\!d^4x\sqrt{-g}\left[M_p^2 R-\!\!\left(\!g^{\alpha\beta}\!-\!\frac{G^{\alpha\beta}}{M_\phi^2}\!\right)\!\!\partial_\alpha\phi\partial_\beta\phi-2V(\phi)\right].
\ee
Note that the coupling scale $M_\phi$ does not receive radiative corrections up to the Planck scale~\cite{sloth}.

The Universe, characterized by the Friedman-Robertson-Walker (FRW) metric 
\be
ds^2=-dt^2+R(t)^2 dx^2\ ,\ee
undergoes a quasi-deSitter expansion during inflation, $\dot R/R\equiv H_I\simeq {\rm const}$. In this case $G^{\alpha\beta}\simeq -3 H^2 g^{\alpha\beta}$, so the kinetic term $\phi$ ceases to be canonical and acquires a normalization factor 
\be
N^2=1+\frac{3H_I^2}{M_\phi^2} \ .
\ee 
The scalar field $\phi$ must then be canonically normalized to 
\be
\bar \phi=N\phi\ ,\ee 
which makes 
\be
V(\phi)\to V(\bar\phi/N)\ .\ee 
In the high friction regime \cite{yuki}, $H_I\gg M_\phi$, the curvatures of the potential are suppressed by factors of $N$, e.g. 
\be
\partial_{\bar\phi}V=V'/N \ll V'\ ,\ee 
where $^\prime$ denotes derivatives w.r.t. $\phi$. Therefore, potentials that are steep in $\phi$ can be flat in $\bar\phi$. 
This is the gravitationally enhanced friction mechanism (GEF) explained in~\cite{yuki} and~\cite{Germani:2011mx}. 

The action \eqref{L} has to be understood as the covariant version of an effective field theory in FRW in the spirit of effective field theory of inflation~\cite{senatore}. Note that after canonical normalization of the inflaton, the actual cut-off (strong-interaction scale) during inflation is $\Lambda\sim(M_p H_I^2)^{1/3}$ \cite{new}. The non-minimal coupling corresponds to the lowest-dimension operator of an expansion in $\Lambda$ .

In these models, the power spectrum of fluctuations and their spectral index are
\be\label{GEF}
P_s=\frac{H_I^2}{8\pi^2 M_p^2}\frac{1}{\epsilon} \quad {\rm and}\quad n_s-1=-8\epsilon+2\eta ,
\ee
where 
\be
\label{GEF2}
\epsilon=\frac{V'^2M_p^2}{2V^2}\frac{1}{N^2}\quad; \quad \eta=\frac{V''M_p^2}{V}\frac{1}{N^2}
\ee 
are the slow-roll parameters of the theory, which satisfy $\eta,\epsilon\ll1$. 
Note, that they are different from the ones in standard minimal inflationary theories.
\vspace{0.1cm}

\section{Axion Dark Matter}\label{sec2}
In the early universe, axion DM is produced around the temperature $T_{\rm osc}$ defined as  
\be
3H(T_\mathrm{osc})\equiv m_a(T_\mathrm{osc})\ee 
where $m_a$ is the axion mass. At this time, the axion starts to oscillate coherently as a condensate of non-relativistic particles. 
The number density of DM axions is 
\be
n_a\simeq \frac{1}{2}m_a a_i^2\ .\ee 
The axion mass is temperature-dependent. Below the QCD phase transition ($T<\Lambda_{\rm QCD}\sim 200$ MeV), $m_a\sim ({77 \rm MeV})^2/f_a$ is fixed by the low energy action, but at $T>\Lambda_{\rm QCD}$  the mass decreases very strongly with temperature. The $T$-dependence has been estimated in the dilute-instanton-gas approximation (see~\cite{Visinelli:2011wa,Bae:2008ue} and refs. therein) and in the interacting-instanton-liquid-model~\cite{Wantz:2009mi}. Albeit fraught with some controversy in the past, the most recent estimates seem to reasonably converge~\cite{Wantz:2009it}.

Assuming radiation domination during the QCD phase transition and standard cosmology afterwards, 
the axion DM abundance today, $\rho_a$, is~\cite{Wantz:2009it} 
\be
\label{axiondmconst}
\frac{\rho^a_{\rm DM}}{\rho^{\rm obs}_{\rm DM}}\simeq \Theta_i^2\begin{cases} 1.7\; \left(\frac{f_a}{10^{12} \rm{GeV}}\right)^{1.184}\quad ; (f_a<f_\Lambda)\\
8\times 10^5 \left(\frac{f_a}{10^{17} \rm{GeV}}\right)^{1.5} \quad ; (f_a>f_\Lambda)\ ,
\end{cases}
\ee
where $f_\Lambda\equiv 3.6\times 10^{17}\rm GeV$ and the observed DM abundance is $\rho^{\rm obs}_{\rm DM}=1.3$ keV/cm$^3$~\cite{planck}. 
The upper expression of \eqref{axiondmconst} corresponds to $T_{\rm osc}>\Lambda_{\rm QCD}$ and the lower to $T_{\rm osc}<\Lambda_{\rm QCD}$. Note that for each value of $f_a\gtrsim 10^{11}$ GeV there is an initial condition for which axions can account for all the DM. We denote it by $\Theta_i(f_a)$ \footnote{We neglect quantum fluctuations in $\Theta_i$, because they are negligible in our model. Corrections to~\eqref{axiondmconst} from anharmonicities are only sizeable for $\Theta_i>1$, i.e. for axion CDM in the $f_a < 5\times 10^{11}$ GeV range. These do not change our conclusions qualitatively and for the sake of the argument we will assume that $f_a > 5\times 10^{11}$ GeV, although we will comment on the low-$f_a$ case later on.}.

The main constraint on this scenario comes from isocurvature perturbations. 
Being essentially massless during inflation,
the axion field receives quantum fluctuations of the order of the Hubble scale, $\delta a\simeq H_I/2\pi\simeq {\rm const}$. Since the axion potential is flat during inflation, these fluctuations will not perturb the total energy density of the universe; such fluctuations are called isocurvature perturbations. 

When the axion mass builds up at $T\simeq \Lambda_{\rm QCD}$ these fluctuations induce non-vanishing density perturbations. Since the relevant curvature perturbations are already at super-horizon scales and therefore frozen, other components have to be perturbed such that the total curvature perturbation vanishes. 
This implies additional temperature perturbations. 
Standard calculations~\cite{Langlois} show that the power spectrum of these perturbations is given by 
\be
P_{iso}=\frac{\delta^{iso}T}{T} \propto - \frac{\delta n_a}{n_a}\ .\ee 
The total power-spectrum is $P(k)=P_{ad}+P_{iso}$ where $P_{ad}$ is the adiabatic component induced by the inflaton. 
Planck observes only adiabatic perturbations, and thus constrains the isocurvature component 
\begin{eqnarray}\label{alphaiso}
\alpha&\equiv& \frac{P_{iso}}{P_{iso}+P_{ad}}\simeq \frac{H_I^2}{P_{s}\pi^2f_a^2\Theta_i^2} 
< 0.039\, (95\% \rm C.L.)
\end{eqnarray}
where $P_s= 2.2\times 10^{-9}$~\cite{planck} and we have assumed $\Theta_i<1$. 

Insisting on axions accounting for all the DM, $\Theta_i(f_a)$ is known, and one finds a widely-discussed~\cite{Wantz:2009it,Raffelt2009} upper bound on $H_I < \pi \sqrt{\alpha P_s} f_a \Theta_i(f_a)$,  
\be\label{HIbound}
H_I < 
\begin{cases}
               2.3\times 10^7 \left(\frac{f_a}{10^{12}\rm GeV}\right)^{0.408}\ {\rm GeV} \quad ; (f_a<f_\Lambda) \\
                3.2\times 10^9\left(\frac{f_a}{10^{17}\rm GeV}\right)^{0.25} \ {\rm GeV}\ \quad ; (f_a>f_\Lambda)  , 
\end{cases} 
\ee 
implying $H_I< 10^{10}$ GeV for $f_a<M_p$.

\section{Saving the Dark Matter Axion}\label{sec3}
\subsection{Small scale Inflation}
%
In order to obtain a small-scale inflationary scenario we consider the model of~\cite{uv}, which is a natural inflation model~\cite{natural} where all scales are sub-Planckian. 
The role of the inflaton is played by a hidden axion (characterized by a decay constant $f_\ha$), which is non-minimally kinetically coupled as in (\ref{L}) and where (\ref{GEF}) and (\ref{GEF2}) hold.
The potential comes from instanton effects of an \emph{extra} hidden gauge group with strong-coupling scale $\Lambda_\ha$. In the small field case ($\ha/f_\ha\ll 1$), one finds 
\be
V(\ha)\simeq\Lambda_\ha^4(2-\frac{\ha^2}{2f_\phi^2})\ee 
which directly implies
\be
\label{1ns}
1-n_s\simeq 2\eta=\frac{M_p^2}{f_\ha^2}\frac{1}{N^2} .  
\ee 
In a minimally coupled scenario, $M_\phi\to \infty$, we have $N=1$ and the Planck measurement $1-n_s\simeq 0.04$ would imply an unreliable new-physics scale $f_\phi \sim 5 M_p$. 
The role of the non-minimal coupling is evident. 
In the high friction regime, $M_\phi\gg H_I$, we have $N\gg1 $ and therefore $f_\phi$ can be easily made sub-Planckian and hence unproblematic. 

The power-spectrum allows to relate $H_I$ to $\ha/f_\ha$, 
\be
H_I=\pi \sqrt{ P_s (1-n_s)/2} \;M_p \frac{\ha}{f_\ha} \simeq 5\times 10^{13}\frac{\ha}{f_\ha}{\rm GeV} . 
\ee
Thus, by considering small $\ha$, we can arbitrarily choose a low $H_I$. Moreover, this is possible for a large range of values of $f_\phi$ by choosing $M_\phi$ to satisfy eq.~\eqref{1ns}. 
In this model it is therefore possible to have axion DM created from the misalignment mechanism~\cite{Wantz:2009it, Raffelt2009} with $H_I\lesssim 10^{10}$ GeV avoiding large isocurvature perturbations as required by~\eqref{HIbound}.
Since the model is single-field, non-Gaussianities are negligible \cite{Maldacena:2002vr}.
This mechanism does not exclusively apply to natural inflation, but to any  model in which inflation takes place close to a maximum of the potential, e.g. Hilltop potentials~\cite{hilltop}.

\subsection{Suppressing isocurvature perturbations}
The isocurvature perturbations are given by the ratio $\delta\Theta/\Theta_i$. 
In order to quantize $\Theta$, we need to canonically normalize it. 
If the axion field is non-minimally coupled to gravity as $\phi$ in \eqref{L}, its kinetic term is
\be
{\cal L}_{a\rm, kinetic}= \frac{1}{2}\left(\!g^{\alpha\beta}\!-\!\frac{G^{\alpha\beta}}{M_a^2}\!\right)\!\!\partial_\alpha a\partial_\beta a\ ,
\ee
with $M_a$ a new energy scale. Note that only a derivative coupling is allowed by the tree-level shift invariance of the axion. During inflation the
canonically-normalized field is $\bar a=N_a a=N_a \Theta f_a$ with 
\be
N_a=1+3 H_I^2/M_a^2\ .\ee 
It is this field that quantum-mechanically produces isocurvature fluctuations during inflation. 
At super-horizon scales the size is 
\be
\langle\delta \bar a^2\rangle =  \frac{H_I^2}{4\pi^2}\ .\ee     
Written in terms of $\Theta$ one has 
 \begin{equation}
\langle \delta \Theta\delta \Theta\rangle = \frac{1}{N_a^2}\frac{H_I^2}{4\pi^2 f_a^2} \ . 
\end{equation}  
Comparing to the minimally coupled case where $N_a=1$ we have
\be
\frac{P_{iso}\Big|_{\rm non-minimal}}{P_{iso}\Big|_{\rm minimal}}\sim \frac{\langle \delta \Theta\delta \Theta\rangle\Big|_{\rm non-minimal}}{\langle \delta \Theta\delta \Theta\rangle\Big|_{\rm minimal}}\simeq N_a^{-2}\ .
\ee
Thus, if $N_a\gg 1$ the isocurvature fluctuations are suppressed with respect to standard expectations. This requires the high friction regime $H_I\gg M_a$. 

Since the power of the isocurvature fluctuations is $1/N_a^2\simeq M_a^2/3 H_I^2$ times smaller in this scenario, the constraint \eqref{alphaiso} turns then into an upper bound on $M_a$
\be
\label{Mabound}
M_a
<M_a^{\rm max}=
\begin{cases}
              4.0\times 10^7 \left(\frac{f_a}{10^{12}\rm GeV}\right)^{0.408}\ {\rm GeV} , \\
                 5.2\times 10^9\left(\frac{f_a}{10^{17}\rm GeV}\right)^{0.25}  \ {\rm GeV} .
\end{cases}
\ee
This implies $M_a< 1.2\times 10^{10} \ {\rm GeV}$ for $f_a< M_p$. 
 Note that the super-horizon evolution of the fluctuations occurs in much the same way as without the non-minimal coupling because the equation of motion 
 \be
 \nabla_\mu \left(\left(g^{\mu\nu}-\frac{G^{\mu\nu}}{M^2_a}\right)\partial_\nu a \right) =0\ee 
still has the trivial solution $a=$const. regardless of the time evolution of $H$.  
 Additionally, since the axionic DM is created during the QCD epoch where $H\ll M_a$, the relic abundance is not influenced by the non-minimal coupling. 

Let us now comment on the range $f_a< 5 \times  10^{11}$ GeV, which requires $\Theta_i>1$. In this regime, our expression for the DM~\eqref{axiondmconst} and $\alpha$~\eqref{alphaiso} are both underestimated~\cite{Kobayashi:2013nva}. The upper limit on $H_I$ has been computed in~\cite{Kobayashi:2013nva}, and it translates directly into our bounds to $M_a^{\rm max}$. 
For $f_a=10^{10.5}(10^{10})$ GeV, which correspond to $\pi-\Theta_i\simeq 10^{-2}(10^{-4})$, the limits are $M_a<10^6(10^3)$ GeV.

We finally note that another mechanism for suppressing isocurvature fluctuations is to force the QCD coupling to be strong during inflation 
by coupling for instance the inflaton to the gluon kinetic term as $(\phi/M_p)^2G_{\mu\nu}G^{\mu\nu}$ with $\phi/M_p\lesssim 1$~\cite{giaaxion}. Interestingly, one can naturally achieve this in the GEF inflation scenario, again pointing to the non-minimal coupling.

\section{Inflation and Dark Matter from the Standard Model}\label{sec4}

We now consider the natural case in which, not only the QCD axion, but also the Higgs boson of the SM is non-minimally kinetically coupled to gravity. This allows the attractive option of producing, with only one mechanism, a successful inflationary scenario and the right abundance of DM.

The Higgs boson ($h$) with the action \eqref{L} (mass scale $\to M_h$) and potential \footnote{Here we assume that, during inflation, $h\gg 246$ GeV, i.e. much larger than the Higgs boson vev for the SM. $h$ is the Higgs component in the unitary gauge.} $V(h)=\frac{\lambda}{4}h^4$ leads to a successful model of inflation, as shown in~\cite{new}. 

The equations \eqref{GEF}, \eqref{GEF2} predict 
\be
H_I=2\pi M_p \sqrt{2 P_s(1-n_s)/5}=9\times 10^{13} {\rm GeV}\ee
but do not fix $M_h$. For this we have to complement them with the Friedman equation (in slow-roll regime)  
\be 
H_I^2\simeq V/3 M_p^2\ ,\ee
leading to
\be
\label{Mh}
M_h=4.0\, M_p (1-n_s)^\frac{5}{4}P_s^\frac{3}{4} \lambda^{-\frac{1}{4}}\ .
\ee
The recent measurement of ATLAS and CMS of the Higgs boson mass, $m_h=126$ GeV,   
give $\lambda=0.26$. However, $\lambda$ runs from the electroweak scale to the $\sim H_I$ scale where our formulas apply. As an order of magnitude estimate of the values of $\lambda$ during inflation, we considered the SM renormalization group equations up to the scale $H_I$, for a recent computation see e.g. \cite{Bezrukov:2012sa,higgsloop}. In order to avoid the electroweak instability problem (see e.g. \cite{ew} and references therein), we consider values of the top mass $m_t\simeq 171 {\rm GeV}$. This is within the $3 \sigma$ range of the measured value. Of course, new physics can in principle have an important role and are a source of uncertainties that we cannot address. The additional non-minimal coupling itself influences the running (at scales $> M_h$). We expect that it softens the running such that $\lambda>0$ even for different values of $m_{t}$, but this important aspect is left for future work \cite{prep}. 
All in all, assuming $\lambda(H_I)>0$ is realized allows to consider $\lambda\sim 0.01$ as an order-of-magnitude estimate. 

If our model is responsible for having Higgs inflation and QCD axions as DM, we would ideally only tolerate a small hierarchy between $M_a$ and $M_h$. 
\begin{figure}[t!]
\includegraphics[width=0.4\textwidth]{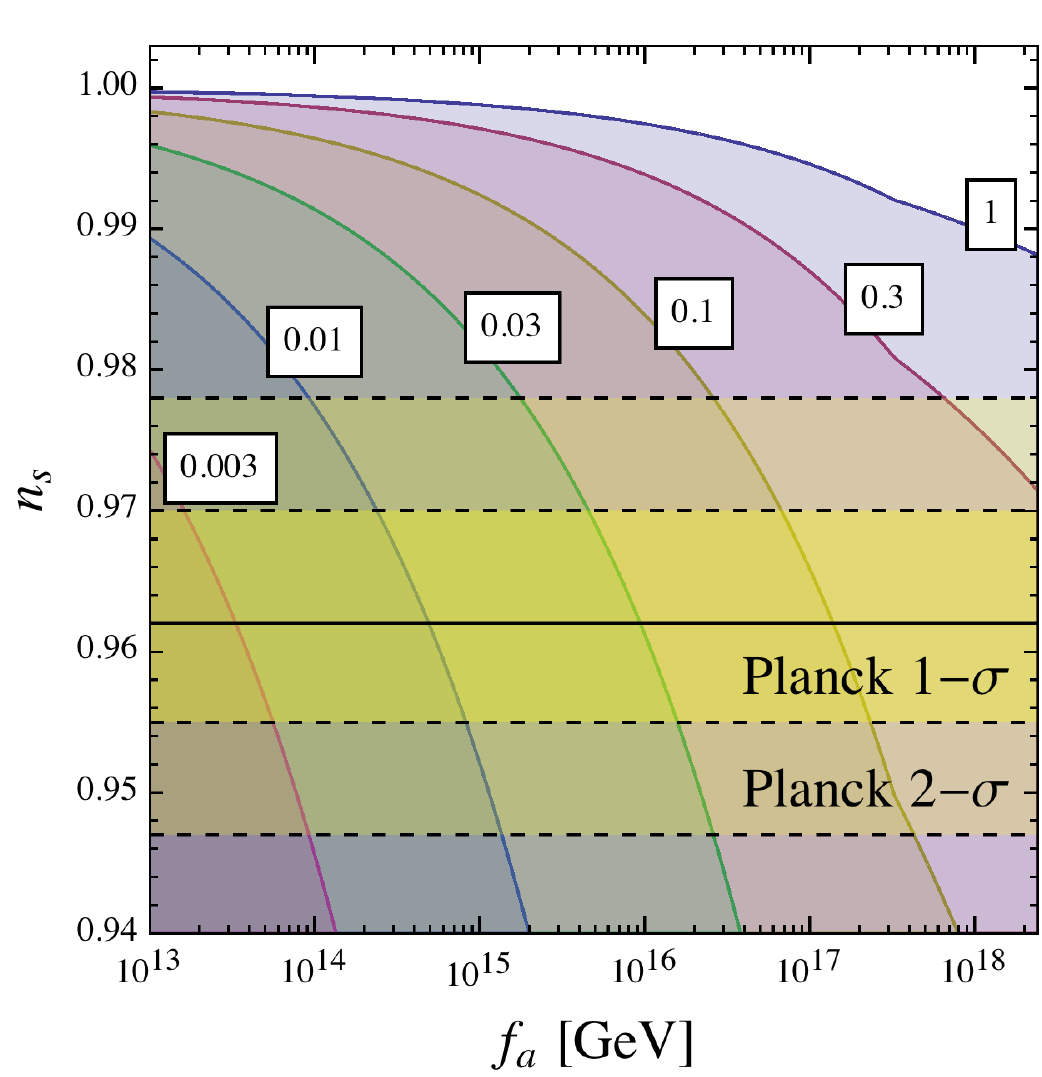}
\caption{
\small Isocontours of $M^{\rm max}_a/M_h$, i.e. the maximum value of $M_a/M_h$ in our model of Higgs inflation and axion DM allowed by Planck constraints on isocurvature perturbations. We have used $\lambda(H_I)=0.01$. The spectral index $n_s$ measured by Planck is shown as yellow bands for 1 and 2$\sigma$ error.  The area outside is disfavored at $95\%$C.L.}
\label{lambdaplot}
\end{figure}
The ratio $M_a/M_h$ is not fixed by our model and Planck data, but it is bounded from above because of the upper limit for $M_a$, c.f.~\eqref{Mabound}. The upper limit $M_a^{\rm max}/M_h$ depends on the value of $f_a$ and generally decreases with decreasing $f_a$. 
In Fig.~\ref{lambdaplot} we show isocontours of $M^{\rm max}_a/M_h$ in the $n_s-f_a$ plane. We see that natural values ($M_a^{\rm max}/M_h \sim 0.1$) are possible for the highest meaningful values of  $f_a\sim M_p$. 
Even for values as small as $f_a\sim 10^{16}$ GeV, we get quite acceptable ratios of 
\be
0.01<M_a^{\rm max}/M_h\lesssim 0.03\ .\ee 
These numbers are relatively sensitive to uncertainties on the measured value of $n_s$ but not on $\lambda$ because it enters mainly through $M_h$, which scales as $(1-n_s)^{5/4}/\lambda^{1/4}$ (see~\eqref{Mh}). Therefore, even though for a complete picture one has to perform the full RG analysis, we are confident that this will not change our main conclusions. 
Note that in this scenario $\Theta_i(f_a)$ is always small and anharmonic effects can be safely neglected.

\section{Conclusions}\label{sec5}

The absence of an indirect observation of gravitational waves by Planck puts a tight, model-independent bound on the Hubble scale during inflation. This bound threats some of the simplest and generic models of inflation such as chaotic inflation with a polynomial potential. However, scenarios with non-minimal kinetic couplings as in~\cite{uv} or~\cite{new} are less constrained because of their small gravitational wave production during inflation~\cite{yuki}. 

The absence of non-Gaussianities and isocurvature perturbations in the CMB challenges instead one of the best motivated dark matter candidates, the QCD axion, if minimally coupled to gravity. 

In this paper we have discussed two possibilities to avoid isocurvature constraints on the axion DM {\it without introducing new degrees of freedom}:

The first one is to consider a consistent low-scale inflationary scenario, as the natural inflationary scenario of \cite{uv}, where all scales are sub-Planckian.

The second mechanism involves instead a modification of how the QCD axion interacts to gravity: The amplitude of isocurvature perturbations is proportional to the (canonical) axion fluctuations normalized to the axion expectation-value today. A non-minimal derivative coupling of the axion to curvatures, ``changes" the canonical normalization of the axion before and after inflation by a factor $N_a$. Where, in first approximation, $N_a$ is the ratio between the Hubble scale during inflation and the coupling constant of the non-minimal interaction $M_\phi$. By choosing $M_\phi$ such to have $N_a\gg 1$, one can easily obtain a largely suppressed spectrum of isocurvature perturbations. Specifically, we showed that the isocurvature constraints of Planck are easily fulfilled with a mild constraint for the coupling constants of the axion-gravity system ($M_\phi$).


As a non trivial final possibility, we have also shown that non-minimal kinetic couplings allow the Higgs boson and the axion to account for both inflation and dark matter, respectively, without introducing a large hierarchy of scales.


\section*{Acknowledgments}
We thank A.~Pritzel, N.~Wintergerst for initial discussions, G.~Dvali, J.~Jaeckel, M.~Goodsell, and G.~Raffelt for reading the first draft {\Sarah and the referees for useful feedback. }
The authors are supported by Humboldt foundation.

\end{document}